\documentclass[12pt]{article}
\usepackage{latexsym}
\usepackage{graphicx}

\catcode`\@=11
\@addtoreset{equation}{section}

\catcode`@=12
\relax

\newcommand{\nn}{\nonumber}

\begin{document}
\thispagestyle{empty}

\renewcommand{\thefootnote}{\fnsymbol{footnote}}
\font\csc=cmcsc10 scaled\magstep1
{\baselineskip=14pt
 \rightline{
 \vbox{\hbox{TIT-HEP-485}
       \hbox{October 2002}
       \hbox{hep-th/0210148}
}}}

\vfill
\begin{center}
{\LARGE\bf
Comments on Effective Superpotentials}
\\
\vspace{3mm}
{\LARGE\bf via Matrix Models
}
\vspace{5mm}
\\

\vfill

{\csc \large  Hiroyuki Fuji}\footnote{
      e-mail address : hfuji@th.phys.titech.ac.jp} and 
{\csc \large Yutaka Ookouchi}\footnote{
      e-mail address : ookouchi@th.phys.titech.ac.jp}\\
\vskip.1in

{\large \baselineskip=15pt
\vskip.1in
  Department of Physics, 
  Tokyo Institute of Technology,\\
  Tokyo 152-8511, Japan
\vskip.1in
}

\end{center}
\vfill

\begin{abstract}
{

Dijkgraaf and Vafa have conjectured that the effective superpotentials
 for $\mathcal{N}=1$ four-dimensional supersymmetric gauge theories can
 be given by the planar diagrams of matrix models. We examine some
 special models with cubic and quartic tree level superpotentials for
 adjoint chiral superfield $\Phi$. We consider the effective
 superpotentials for the classical vacuum $\Phi=0$ for $U(N)$ and
 $SO(N)/Sp(N)$ gauge theories. 
We evaluate the effective superpotentials exactly in terms of the matrix model and in
 terms of closed string theory on Calabi-Yau geometry with fluxes. As a
 result we find their perfect agreements. 

}
\end{abstract}
\vfill

\setcounter{footnote}{0}
\renewcommand{\thefootnote}{\arabic{footnote}}
\newpage
\setcounter{page}{1}


\section{Introduction}

Since 't Hooft introduced an idea \cite{Hooft} of the correspondence
between the large $N$ gauge theory with certain string theory, many
dualities had been discussed with this idea. One example of these
dualities is the relation between Chern-Simons gauge theory and A-type
topological strings \cite{GV,SV,OV,AMV}. This duality between the
topological theories can be embedded in type IIA/IIB superstring
\cite{VAFA}. The discussion for type IIB superstrings was extended in
\cite{CIV,CKV,CFIKV,OH,FO,DOT,CV}. The equivalence of effective
superpotential for ${\cal N}=1$ gauge theory with that of type IIB
superstring theory on Calabi-Yau manifold with fluxes \cite{Gukov,TV} was proved in \cite{CV}.

Recently Dijkgraaf and Vafa conjectured the correspondence between
${\cal N}=1$ gauge theories with matrix models based on the reduction 
of B-type topological string theory with B-branes \cite{DV1}.
They proposed the effective superpotentials for a large class of
$\mathcal{N}=1$ super Yang-Mills theories are computed from matrix
models \cite{DV1,DV2,DV3}. In particular the effective superpotentials
for the gauge theories
are given by the genus zero free energies which are the summations of
planar diagrams in matrix models. This perturbative sum of planar
diagrams turns out to be a nonperturbative sum over fractional
instantons in the ${\cal N}=1$ gauge theory. For $\mathcal{N}=1^*$ theory they found the coincidence with those of field theory results. In \cite{Dorey1,Dorey2} Dijkgraaf-Vafa's proposal was tested further for a family of deformations of $\mathcal{N}=4$ super Yang-Mills theory involving an arbitrary polynomial superpotential for one of the three adjoint chiral superfields.

It is important to check this proposal 
for ${\cal N}=1$ super Yang-Mills theory which is the deformed 
${\cal N}=2$ super Yang-Mills theory by the superpotential for 
adjoint chiral superfields.
In particular
the models with cubic and quartic superpotentials are quite interesting,
because the matrix model corresponding to these potentials have been
studied well and the effective superpotential for the gauge theory was
explicitly computed up to fourth order in glueball superfields from the 
dual closed string theory on Calabi-Yau manifold with RR fluxes \cite{CIV,FO}. Therefore these models are appropriate for the check of the Dijkgraaf-Vafa conjecture for $\mathcal{N}=1$ theories.

The elegant proof of this correspondence for ${\cal N}=1$ super
Yang-Mills theory is given in \cite{CM}.
It is also very important to find the effective superpotential directly
for this case and check this correspondence explicitly.
 In this note we explicitly compute the effective superpotentials for
 $U(N)$ and $SO(N)/Sp(N)$ gauge theories with cubic/quartic tree level
 superpotentials around the classical vacuum $\Phi=0$ in the
 context of matrix model and the context of Calabi-Yau manifold with fluxes. The condition $\Phi=0$ means the gauge symmetry remain unbroken.
While preparing the manuscript for submission, 
we recieved \cite{Fe} which overlaps with $U(N)$ gauge theory result in our work.

The organization of this paper is as follows: In section 2, we discuss
the effective superpotential for the $U(N)$ gauge theory with cubic tree
level superpotential. At first we review the Calabi-Yau manifold which
is the dual of the gauge theory and compute the effective superpotential
around the classical vacuum $\Phi=0$. Next we review the derivation of
 genus zero free energy of the Hermitean matrix model and compute its derivative
 with respect to $S=g_sN$. Comparing these results,
we can check the Dijkgraaf-Vafa conjecture exactly for this case. In section 3, we discuss the effective superpotential for the $SO(N)/Sp(N)$ gauge theory with quartic tree level superpotential as in section 2.  
 
%

\section{$U(N)$ gauge theory with the cubic superpotential}
In this section we discuss the effective superpotential for the $U(N)$
gauge theory with cubic tree level superpotential around the classical
vacuum $\Phi=0$. Since the equivalence of the effective superpotential
for the gauge theory with that of type IIB superstring theory on Calabi-Yau manifold with fluxes \cite{Gukov,TV} was proved in \cite{CV}, we concentrate on the Calabi-Yau manifold with fluxes.
At first we calculate the effective superpotential from the Calabi-Yau
manifold with fluxes. In the classical vacuum $\Phi=0$, since there is
only one period of the Calabi-Yau manifold, which is identified with glueball superfield, we can obtain the exact result. Next we consider the effective superpotential from matrix model context in terms of the Dijkgraaf-Vafa conjecture. The Hermitean matrix model with cubic action is also exactly solvable, therefore we can obtain the genus zero free energy exactly.  Then we discuss the Dijkgraaf-Vafa conjecture with these results.

\subsection{The analysis for the Calabi-Yau geometry with fluxes}
We begin with the review of the effective superpotential from the
Calabi-Yau manifold with fluxes \cite{CIV}. We consider $4d$
$\mathcal{N}=1$, $U(N)$ super Yang-Mills theory with tree level superpotential for adjoint chiral superfield $\Phi$
\begin{eqnarray}
W_{\mathrm{tree}}=\frac{m}{2}\mathrm{Tr}\Phi^2+\frac{g}{3}\mathrm{Tr}\Phi^3. \label{treeU}
\end{eqnarray}
Since the eigenvalues of $\Phi$ are given by the roots of $W^{\prime}_{\mathrm{tree}}=0$, the gauge group is broken as 
\begin{eqnarray}
U(N) \to U(N_1)\times U(N_2),\ \ \ \ \ N_1+N_2=N.
\end{eqnarray}
The dual description of the gauge theory in this vacuum  \cite{CIV} was
given \ by the following deformed Calabi-Yau manifold with RR-flux $H^{RR}$ and NSNS B-field $H^{NS}$,
\begin{eqnarray}
{W_{\mathrm{tree}}^{\prime}}^2(x)+y^2+z^2+w^2+f_1(x)=0.\label{CY}
\end{eqnarray}
where $W_{\mathrm{tree}}$ is the tree level superpotential for the gauge theory and $f_1$ is a degree one polynomial. In this Calabi-Yau manifold we can form an integral basis of 3-cycles, $A_i$ and $B_i$, which form a symplectic paring 
\begin{eqnarray}
(A_i,B_j)=-(B_j,A_i)=\delta_{ij},\ \ \ (A_i,A_j)=(B_i,B_j)=0,\ \ i,j=1,2
\end{eqnarray}
The above geometry have $N_i$ units of RR-flux through the $A_i$ cycle, and an NS-flux $\alpha$ through each of the dual noncompact $B_i$ cycles. $\alpha$ is identified with the complex bare gauge coupling of the gauge theory. 
The periods of the Calabi-Yau manifold is given by the integral of the holomorphic 3-form $\Omega$ over these cycles. Since $B_i$-cycles are noncompact cycle, we need the cut-off $\Lambda_0$ for the integral of $B_i$-cycle
\begin{eqnarray}
\int_{A_i} \Omega \equiv S_i,\ \ \ \ \int_{B_i}^{\Lambda_0}\Omega \equiv \Pi_i.
\end{eqnarray}
The period integrals over $A_i$ cycle are identified with the glueball superfield $S=\frac{1}{32\pi}\mathrm{Tr}W_{\alpha}W^{\alpha}$  whose  lowest component is the gluino bilinear. 
These period integrals of the holomorphic 3-form can be written as integrals of the effective one-form $\omega$ over projections of the cycles to the $x$-plane. The one-form $\omega$ is given by doing the $\Omega$ integral over the $S^2$ cycle corresponding to the $y,z,w$ coordinates
\begin{eqnarray}
S_i=\frac{1}{2\pi i}\int \omega,\ \ \ \Pi_i=\frac{1}{2\pi i}\int^{\Lambda_0} \omega ,\ \ \ \  \omega=dx \sqrt{{W^{\prime}}^2(x)+f_{n-1}(x)}.
\end{eqnarray}

We want to consider the special vacuum in which the gauge symmetry is
unbroken $U(N)\to U(N)$ \footnote{In the dual gauge theory, this case
correspond to all of $N$ D5-branes wrap only on the blown-up
$\mathbf{S^2}$ at the origin.}.
In this classical vacuum there is only one period and the $x$-plane have one brunch-cut. In this case the effective one form is written as
\begin{eqnarray}
dx \sqrt{{W^{\prime}}^2(x)+f_1(x)}=dx g (x+\Delta+x_3)\sqrt{(x-2x_1)(x-2x_2)}.
\end{eqnarray}
where $\Delta=m/g$. Comparing the coefficients of $x^3$ and $x^2$, we get the following relations
\begin{eqnarray}
x_3=x_1+x_2, \qquad (x_1-x_2)^2+2(x_1+x_2) [\Delta+x_1+x_2]=0. \label{CYrelation}
\end{eqnarray}
It is also convenient to introduce new variables given by
$\bar{\sigma} \equiv x_1+x_2,\mu \equiv x_1-x_2$.
Using these variables, we can rewrite the latter relation of (\ref{CYrelation})
\begin{eqnarray}
\mu^2+2\bar{\sigma}[\Delta+\bar{\sigma}]=0. \label{CYconst}
\end{eqnarray}
The period integral is explicitly evaluated,
\begin{eqnarray}
\frac{S}{g}=\frac{1}{2\pi i}\int_{2x_1}^{2x_2}dx (x+\Delta+\bar{\sigma})\sqrt{(x-2x_1)(x-2x_2)}=\frac{\Delta+2\bar{\sigma}}{4}\mu^2, \label{periodS}
\end{eqnarray} 
Using this result (\ref{periodS}), we rewrite the constraint (\ref{CYconst}) 
\begin{eqnarray}
2\frac{S}{g}+\bar{\sigma} (\Delta+2\bar{\sigma})(\bar{\sigma}+\Delta)=0. \label{UCYcons}
\end{eqnarray}
$\bar{\sigma}$ is solved by a power series in S around $\bar{\sigma}=0$,
\begin{eqnarray}
\bar{\sigma}=-\frac{\Delta}{4}\sum_{k=1}^{\infty}\left(\frac{8 S}{g \Delta^3} \right)^k\frac{\Gamma\left((3k-1)/2 \right)}{\Gamma\left((k+1)/2 \right)}. \label{UCYpower}
\end{eqnarray}
The dual period is written as the function of $\sigma$ which have the constraint (\ref{UCYcons})
\begin{eqnarray}
\frac{2\pi i}{g}\Pi&=&\int^{\Lambda_0}_{2x_2}(x+\bar{\sigma}+\Delta)\sqrt{(x-2x_1)(x-2x_2)}dx  \nn \\
{}&=& \frac{1}{g}W(\Lambda_0)-\frac{\bar{\sigma}^2}{6}(2\bar{\sigma}+3\Delta)+\frac{S}{g}\left[\log \Lambda_0+\log \Delta-\log (2\bar{\sigma}+\Delta) \right]. \label{dualperiod}
\end{eqnarray}
The effective superpotential generated from fluxes is given \cite{TV} by
\begin{eqnarray}
W_{\mathrm{eff}}&=&\int \Omega \wedge (H^{RR}+\tau H^{NS})=2\pi i(N\Pi- \tau S) \nn \\
{}&=&   NS \log \left(\frac{S}{\Lambda_0^3}\right)-2\pi i \tau S+ -\frac{Ng \sigma^2}{6}(2\bar{\sigma}+3\Delta) \nn \\
 {}&& \qquad \qquad+NS\left[\log \Lambda_0+\log \Delta-\log (2\bar{\sigma}+\Delta) \right] \label{Uexactre}
\end{eqnarray}
where we ignored the irrelevant constant $W(\Lambda_0)$. Thus we found the effective superpotential exactly. Substituting the power series result (\ref{UCYpower}) to the exact result (\ref{Uexactre}), we obtain
\begin{eqnarray}
W_{\mathrm{eff}}(S)\!\! &=& \!\! NS \log \left(\frac{S}{\Lambda_0^3}\right)-2\pi i \tau S+ \frac{N}{2}\sum_{k=1}^{\infty}\left(\frac{8g^2}{m^3}\right)^k \frac{S^{k+1}\Gamma(3k/2)}{(k+1)! \ \Gamma(k/2+1)} \nn \\
{}&=& \!\!  \cdots+N \left[2S^2+\frac{32}{2}S^3+\frac{280}{3}S^4+\cdots \right]. \label{UCYresult}
\end{eqnarray}
The coefficients of first three terms agree with those of the result already known in \cite{CIV}.

\subsection{Hermitean matrix model analysis}
The matrix model describing the U(N) gauge theory with tree level superpotential (\ref{treeU}) is discussed in $\cite{DV1}$. 
For this matrix model the partition function is given by
\begin{eqnarray}
Z=\frac{1}{\mathrm{Vol}\left(U(N) \right)}\int d \Phi \cdot \mathrm{exp}\left(-\frac{1}{g_s} \mathrm{Tr}W_{\rm{tree}}(\Phi) \right), \label{Upartition}
\end{eqnarray}
where $\Phi$ is a $N\times N$ Hermitean matrix. This partition function
is evaluated perturbatively by double line Feynman diagrams with orientation and has a consistent saddle-point approximation of the form
\begin{eqnarray}
Z=\mathrm{exp}\sum_{g\ge 0}g_s^{2g-2}\mathcal{F}_g(S),
\end{eqnarray}
in the 't Hooft limit $N\gg 1$ and $g_s \ll 1$, while keeping finite the
't Hooft coupling $S=g_s N$ \cite{DV1}. $\mathcal{F}_g$ is given by the
sum of all diagrams with  genus $g$. As discussed in ordinary quantum field theory, the exponentiation of the connected diagrams generates all diagrams, so the free energy contributes only from connected diagrams. 

Since the eigenvalues $\lambda_i$ $(1\le i\le N)$ of $\Phi$ become continuous in the large $N$ limit, let us introduce a function  $\lambda(x)$ such that $\lambda_i=\lambda (i/N)$.
Then  the genus zero free energy of the model may be given by  the integral,
\begin{eqnarray}
\mathcal{F}_0=-\frac{g_s^2N^2}{g_s}\int_0^1 dx \left(\frac{m}{2N}\lambda^2(x)+\frac{g}{3N}\lambda^3(x) \right)+g_s^2 N^2\int_0^1 \int_0^1 dx dy \log |\lambda (x) -\lambda (y)|.
\end{eqnarray}
The second term comes from the Vandermonde determinant corresponding to the Jacobian. The details of the computation of this genus zero free energy have been discussed in \cite{Zuber}. 

First of all we introduce the density of eigenvalues $\rho(\lambda)$ defined by
\begin{eqnarray}
2\frac{dx}{d\lambda}=\rho(\lambda),\ \ \ \ \int_{2a}^{2b} d\lambda \, \rho(\lambda)=2. \label{Urhodef}
\end{eqnarray}
Outside the interval $(2a,2b)$ the value is zero. Using this quantity,
we obtain  the equation of motion in the large $N$ limit 
\begin{eqnarray}
\frac{m}{g_sN}\lambda +\frac{g}{g_s N}\lambda^2=\int_{2a}^{2b} \!\!\!\!\!\!\!\!\!\!- \ \ \ d\mu \frac{\rho(\mu)}{\lambda-\mu}
\end{eqnarray}
where the bar of the integral means the principal part. The solution is obtained by introducing the resolvent,
\begin{eqnarray}
\omega(\lambda)=\int_{2a}^{2b}d\mu \frac{\rho(\mu)}{\lambda-\mu}
\end{eqnarray}
where complex $\lambda$ is defined outside of the real interval $(2a,2b)$. It behaves as $2/\lambda$ when $|\lambda|$ goes to infinity. The factor $2$ comes from the definition (\ref{Urhodef}). When $\lambda$ approaches the interval $(2a,2b)$, it behaves as
\begin{eqnarray}
\omega(\lambda\pm i0)=\frac{m}{g_s N}\lambda +\frac{g}{g_s N}\lambda^2 \mp i\pi \rho(\lambda).
\end{eqnarray}
There is a unique function, which satisfies these requirements, 
\begin{eqnarray}
\omega(\lambda)=\frac{m}{g_s N}\lambda+\frac{g}{g_s N}\lambda^2-\left(\frac{m}{g_sN}+\frac{g}{g_s N}(a+b)+\frac{g}{g_s N}\lambda \right)\sqrt{(\lambda-2a)(\lambda-2b)},
\end{eqnarray}
with the following constrains,
\begin{eqnarray}
  0\!\!\! &=&\!\!\! 2(a+b)\left[m+g(a+b) \right]+g(b-a)^2 \nn \\
    4g_sN \!\!\!&=&\!\!\! (b-a)^2\left[m+2g(a+b) \right].
   \label{const}
\end{eqnarray}
Since only the singular part contributes to the density of eigenvalues, we obtain the density of eigenvalues,
\begin{eqnarray}
\rho(\lambda)&=&\frac{1}{2\pi i}\left(\omega(\lambda+i0)-\omega(\lambda-i0) \right) \nn \\
&=&  \frac{1}{\pi}\left(\frac{m}{g_s N}+\frac{g}{g_s N}(a+b)+\frac{g}{g_s N}\lambda^2 \right)\sqrt{(\lambda-2a)(2b-\lambda)}. \label{Urho}
\end{eqnarray}
It is convenient to introduce the single parameter $\sigma=\frac{g}{m} (a+b)$.
Using this parameter, we may rewrite the constraints (\ref{const}) as 
\begin{eqnarray}
\frac{2g^2Ng_s}{m^3}+\sigma(1+\sigma)(1+2\sigma)=0. \label{Uconst2}
\end{eqnarray}
The expansion of $\sigma$ as a power series in $g$ is given by
\begin{eqnarray}
\sigma=-\frac{1}{4}\sum_{k=1}^{\infty} \left(\frac{8g^2Ng_s}{m^3}\right)^k\frac{\Gamma((3k-1)/2)}{\Gamma((k+1)/2)}. \label{Usol}
\end{eqnarray}
We rewrite the genus zero free energy in terms of these quantities as
\begin{eqnarray}
\mathcal{F}_0&=&-\frac{g_s^2N^2}{2g_s}\int_{2a}^{2b} d \lambda \rho(\lambda) \left(\frac{m}{2N}\lambda^2(x)+\frac{g}{3N}\lambda^3(x) \right) \nn \\
{}&&\qquad \qquad+ \frac{g_s^2 N^2}{4}\int_{2a}^{2b} \int_{2a}^{2b} d\lambda d\mu \rho(\lambda)\rho(\mu) \log |\lambda -\mu|.
\end{eqnarray}
With the explicit expression (\ref{Urho}), we can integrate out the $\lambda$.
\begin{eqnarray}
\mathcal{F}_0=\frac{g_s^2N^2}{3}\frac{\sigma(3\sigma^2+6\sigma+2)}{(1+\sigma)(1+2\sigma)^2}-\frac{g_s^2N^2}{2}\log (1+2\sigma). 
\end{eqnarray}
This is the exact result of genus zero free energy.
Using the relation (\ref{Uconst2}),(\ref{Usol}) we get the genus zero free energy as a power series in $g_sN$,
\begin{eqnarray}
\mathcal{F}_0=\frac{g_s^2 N^2}{2}\sum_{k=1}^{\infty}(\frac{8g^2g_sN}{m^3})^k\frac{\Gamma(3k/2)}{(k+2)! \ \Gamma(k/2+1)}. \label{Uresult}
\end{eqnarray}
For the later discussion for the examination of Dijkgraaf-Vafa conjecture, we calculate the derivative of the genus zero free energy with respect to 't Hooft coupling $S\equiv g_sN$.
\begin{eqnarray}
\frac{\partial \mathcal{F}_0}{\partial S}&=&\frac{2S}{3}\frac{\sigma(3\sigma^2+6\sigma+2)}{(1+\sigma)^2(1+2\sigma)^2}-S \log (1+2\sigma)\nn \\
&&+S^2\left(\frac{\partial \sigma}{\partial S}\right)\frac{3+14\sigma+20\sigma^2+12\sigma^3+4\sigma^4}{(1+\sigma)^2(1+2\sigma)^3}.
\end{eqnarray}
Using the derivative of constraint (\ref{Uconst2}) with respect to S
\begin{eqnarray}
\frac{\partial \sigma}{\partial S}=-\frac{2g^2}{m^3}\frac{1}{1+6\sigma+6\sigma^2},
\end{eqnarray}
we obtain
\begin{eqnarray}
\frac{\partial \mathcal{F}_0}{\partial S}=-\frac{m^3}{6g^2}\sigma^2 (3+2\sigma)-S \log (1+2\sigma).
\end{eqnarray}
Since $\bar{\sigma}=\Delta\sigma$, 
this result agree with
the fractional instanton contribution part of 
 the dual period integral $\Pi$ (\ref{dualperiod}) in the previous subsection.

\subsection{The check of Dijkgraaf-Vafa conjecture for $U(N)$ case}

In \cite{TV} the effective superpotential is given by the period $S$ of
Calabi-Yau manifold (\ref{CY})
\begin{eqnarray}
W_{\mathrm{eff}}(S)= NS \log \left(\frac{S}{\Lambda_0^3}\right)-2\pi i \tau S+N\frac{\partial F_{\mathrm{pert}}(S)}{\partial S} , \ \ \ F_{\mathrm{pert}}(S)=\sum_{i}c_i S^i
\end{eqnarray}
where $F_{\mathrm{pert}}(S)$ is the prepotential for the special
 geometry of (\ref{CY}) and expanded perturbatively in terms of $S$.
This superpotential corresponds to that of gauge theory in the
 confining phase by identifying the periods $S$ with 
the glueball superfield $S=\frac{1}{32\pi^2}\mathrm{Tr}W_{\alpha}W^{\alpha}$. 
Under this identification, 
the first two terms correspond to the Veneziano-Yankielowicz form \cite{VY} and $\tau$ is the bare complex coupling of the gauge theory.

In \cite{DV3} it is conjectured that 
$F_{\mathrm{pert}}(S)$  equals to 
the free energy $\mathcal{F}_0(g_s)$ of the matrix model 
under the identification between the glueball superfield $S$ with
the 't Hooft coupling $g_s N$.
\begin{eqnarray}
W_{\mathrm{eff}}(S)&=& NS \log \left(\frac{S}{\Lambda_0^3}\right)-2\pi i \tau S+N\frac{\partial \mathcal{F}_0}{\partial S} \nn \\
{}&=& NS \log \left(\frac{S}{\Lambda_0^3} \right)-2\pi i \tau S+N\left[\frac{3m^3}{2g^2}\sigma^2 (3+2\sigma)+S \log (1+2\sigma)\right]
\end{eqnarray}
where the first two terms come from the volume factor in
(\ref{Upartition}). The discussion of this volume factor is discussed in 
\cite{OV,DV1,Mor}. This effective superpotential agrees exactly with the result (\ref{Uexactre}) from the Calabi-Yau manifold with fluxes.
Furthermore, using the result (\ref{Uresult}) we have easily got the exact effective superpotential for the gauge theory with the tree level superpotential of (\ref{treeU})
\begin{eqnarray}
W_{\mathrm{eff}}(S)\!\! &=& \!\! NS \log \left(\frac{S}{\Lambda_0^3}\right)-2\pi i \tau S+ \frac{N}{2}\sum_{k=1}^{\infty}\left(\frac{8g^2}{m^3}\right)^k \frac{S^{k+1}\Gamma(3k/2)}{(k+1)! \ \Gamma(k/2+1)} \nn \\
{}&=& \!\!  \cdots+N \left[2S^2+\frac{32}{2}S^3+\frac{280}{3}S^4+\cdots \right].\label{Ueff}
\end{eqnarray}
This is the exact effective superpotential summed up all instanton
 corrections and agrees perfectly with (\ref{UCYresult}).
Thus we find that $F_{\mathrm{pert}}(S)=\mathcal{F}_0(g_s)$ holds exactly. 
Therefore
Dijkgraaf-Vafa's conjecture is checked 
for the special vacuum $\Phi=0$.

\section{$SO(N)/Sp(N)$ gauge theories with quartic superpotential}

In this section we discuss the effective superpotential for
$SO(N)/Sp(N)$ gauge theories with quartic tree level superpotential around
the classical vacuum $\Phi=0$. We proceed the discussion in the same way
as section 3. But there are some differences. The geometry dual to the
gauge theory has $Z_2$ identification. The matrix models corresponding to
the $SO(N)/Sp(N)$ gauge theories are not Hermitean but real symmetric and 
self-dual quaternionic respectively.

\subsection{The analysis for the Calabi-Yau geometry with fluxes}
We begin with somewhat general discussion for the $SO/Sp$ gauge theories for the later discussion of matrix model.
 In order to realize $SO(N)/Sp(N)$ gauge theories, we need introduce orientifold-plane. When an orientifold-plane is introduced, the Calabi-Yau manifold must be invariant under the complex conjugation \cite{SV,OH}
 \begin{eqnarray}
(x,y,z,w) \to (\bar{x},\bar{y},\bar{z},\bar{w}). 
 \end{eqnarray}
 Then we can only realize an $SO/Sp$ gauge theories with tree level superpotential of even function. The large $N$ dual of this theory is found via geometric transition \cite{OH,FO}. Through the geometric transition, the dual geometry for the $SO(N)/Sp(N)$ gauge theories is given by
\begin{eqnarray}
{W_{\mathrm{tree}}^{\prime}}^2(x)+y^2+z^2+w^2+f_{2n-2}(x)=0,\ \ \ W_{\mathrm{tree}}=\sum_{k=1}^{n+1}\frac{g_{2k}}{2k}x^{2k}  \label{SOgeometry}
\end{eqnarray}
where $W_{\mathrm{tree}}$ is the tree level superpotential for the gauge theory and $f_{2n-2}$ is a degree $2n-2$  polynomial.

In the following we concentrate on the case with the quartic tree level superpotential,
\begin{eqnarray}
W_{\mathrm{tree}}=\frac{m}{2}\mathrm{Tr} \Phi^2+\frac{g}{4}\mathrm{Tr} \Phi^4. \label{SOfourth}
\end{eqnarray}
As in the case of $U(N)$ gauge theory, we can define an integral basis of 3-cycles which form a symplectic paring and the periods of the Calabi-Yau manifold. In this case since we also consider the special classical vacuum $\Phi=0$, there is only one period and the dual period and the $x$-plane also have one brunch-cut. We write the effective one form as follows
\begin{eqnarray}
2dx \sqrt{{W^{\prime}}^2+f_2}=2dx g (x^2+\Delta+x_3)\sqrt{(x-2\mu)(x+2\mu)}.
\end{eqnarray}
where $\Delta=m/g$. From the coefficient of $x^4$ we get the following relations, $x_3=2\mu^2$.
The period integral is explicitly evaluated,
\begin{eqnarray}
\frac{S}{g}=\frac{1}{\pi i}\int_{-2\mu}^{2\mu}dx (x^2+\Delta+2\mu^2)\sqrt{4\mu^4-x^2}=6\mu^4+2\Delta \mu^2. \label{periodSO}
\end{eqnarray}
In terms of $S/g$, $\mu^2$ can be solved 
\begin{eqnarray}
\mu^2=\frac{-\Delta+\sqrt{\Delta^2+6S/g}}{6}.\nonumber
\end{eqnarray}
The dual period is evaluated as
\begin{eqnarray}
\frac{\pi i}{g}\Pi&=&\int_{2\mu}^{\Lambda_0}dx (x^2+\Delta+2\mu^2)\sqrt{4\mu^4-x^2} \nn \\
{}&=& \frac{1}{g}W(\Lambda_0)-\frac{S}{g}\log \Lambda_0 +\frac{S}{2g}\log \frac{S}{2g}-\frac{S}{2g}-\frac{1}{2}\left(\Delta{\mu}^2-\frac{S}{2g}\right)+\frac{S}{2g}\log {\mu}^2. \label{SOdualperi}
\end{eqnarray}
The effective superpotential generated from fluxes \cite{OH} is given by
\begin{eqnarray}
W_{\mathrm{eff}}&=&\int \Omega \wedge (H^{RR}+\tau H^{NS})=2\pi i\left[\left(N+2 \right)\Pi-\tau S\right] \nn \\
{}&=&   \left(N\mp 2 \right)S \log \left(\frac{S}{\Lambda_0^3}\right)-2\pi i \tau S \nn \\
{}&&\quad +(N\mp 2)\left[-2S\log \Lambda_0 +S\log \frac{S}{g}+S-\frac{S}{2}({\mu}^2-1)+S\log {\mu}^2\right] \label{SOCYexact}
\end{eqnarray}
where we ignored the irrelevant constant $W(\Lambda_0)$. 
In the above expression, upper signs are taken for $SO(N)$ gauge theory
and lower signs are taken for $Sp(N)$ gauge theory.
We obtained the effective superpotential for the gauge theory with tree level superpotential (\ref{SOfourth}). From the relation (\ref{periodSO}) $\mu$ is written as a function of $S$. Substituting to the effective superpotential, we obtain
\begin{eqnarray}
W_{\mathrm{eff}}(S) \!\! &=& \!\!  \hat{N}S \log \left(\frac{S}{\Lambda_0^3}\right)-2\pi i \tau S-\hat{N} \sum_{k=1}^{\infty} \left(-\frac{3g }{2m^2} \right)^k \frac{S^{k+1}}{2} \frac{(2k-1)!}{k!(k+1)!} \nn \\{} \!\! &=& \!\!  \cdots+\hat{N} \left[\frac{3}{2}\left(\frac{g}{4m^2} \right)S^2-\frac{9}{2} \left(\frac{g^2}{8m^4} \right)S^3+\frac{45}{2} \left(\frac{g^3}{16m^6} \right)S^4+\cdots \right], \label{SOeff}
\end{eqnarray}
where $\hat{N}=N\mp 2$.
The coefficients of the first three terms agree with those of our
result in our previous work \cite{FO}.

\subsection{Matrix model analysis}
First we discuss somewhat general properties of real symmetric 
matrix model and 
self-dual quaternionic matrix model with the potential,
\begin{eqnarray}
W=\sum_{k=0}^{n}u_{2k}{\Phi}^{2k}.  \label{SOtree}
\end{eqnarray}
In this case the partition function of the corresponding matrix models is given by 
\begin{eqnarray}
Z=\frac{1}{\mathrm{Vol}\left(G\right)}\int d \Phi \cdot \mathrm{exp}\left(-\frac{1}{g_s} \mathrm{Tr}W(\Phi) \right)
\end{eqnarray}
where 
 $\Phi$ is $N\times N$ real symmetric matrix for $G=SO(N)$ and 
$N\times N$ self-dual quaternionic matrix for $G=Sp(N)$. 
For these matrix models, the Feynman diagrams become unoriented double
 line diagrams. So the large $N$-limit of this model may be described in terms of a simple $\Phi^4$-theory with single lines in which all non-planar diagrams are omitted. 

In this case the analysis which is discussed for Hermitean matrix model
in \cite{DV1} is slightly changed. 
In fact the power of the Vandermonde determinant $\Delta$ is not squared 
and the partition functions become\cite{Me,FGZ} 
\begin{eqnarray}
&&Z=\int \prod_{i=1}^N d\lambda_i \cdot \Delta(\lambda)\cdot \mathrm{exp}\left(-\frac{1}{g_s}\sum_{i=1}^NW(\lambda_i) \right) \quad {\rm for }\; SO(N),\nonumber \\
&&Z=\int \prod_{i=1}^{N/2} d\lambda_i \cdot \Delta(\lambda)^4\cdot \mathrm{exp}\left(-\frac{1}{g_s}\sum_{i=1}^{N/2} 2W(\lambda_i) \right) \quad {\rm for }\; Sp(N)
\end{eqnarray}
where $N$ is even for $Sp(N)$.

We proceed the discussion of loop equation as in
\cite{DV1}.\footnote{Here we will concentrate on the $SO(N)$ case. 
For the $Sp(N)$ case there is slight change of the coefficient in
(\ref{loop}), but they can be
absorbed in the normalizations of $f(x)$ and $y(x)$.}
In the large $N$ limit the loop equation becomes
\begin{eqnarray}
\omega^2(x)+\frac{2}{\mu}\omega (x)W^{\prime}(x)+\frac{1}{\mu^2}f(x)=0 \label{loop}
\end{eqnarray}
where $\mu=g_s N$ is the 't Hooft coupling, $\omega$ is the resolvent and $f(x)$ is given by
\begin{eqnarray}
f(x)=\frac{2\mu}{N}\sum_i\frac{W^{\prime}(x)-W^{\prime}(\lambda_i)}{x-\lambda_i}. \label{SOdiff}
\end{eqnarray}
We have an algebraic loop equation because of the large $N$ limit. Let us introduce a new variable,
\begin{eqnarray}
y(x)=\mu \, \omega(x)+W^{\prime}(x).
\end{eqnarray}
Using this variable we can rewrite the loop equation (\ref{loop}) as
\begin{eqnarray}
y^2-W^{\prime}(x)^2+f(x)=0.
\end{eqnarray}
Thus we obtain a (singular) hyperelliptic curve in the $(x,y)$-space. The remaining
analysis is  the same as in \cite{DV1}. 
In the following, we will evaluate the derivative of the free energy
${\cal F}_0(\mu)\equiv g_s^2\log Z$ in terms of the 't Hooft coupling $\mu$.

Let us remark about the hyperelliptic curve. As in \cite{DV1,BEH} from the relation (\ref{SOdiff}) we can obtain the coefficient of $f(x)=\sum_{k=0}^{n-1} 2\mu b_k x^k$ explicitly,
\begin{eqnarray}
b_{2k}=(2k+2)u_{2k+2}+\sum_{s=1}^{n+1}(2k+2s)u_{2k+2s}\frac{\partial \mathcal{F}_0}{\partial u_{2s-2}}\ , \ \ b_{2k+1}=0, \ \ k=1,\cdots ,\frac{n-1}{2}  
\end{eqnarray}
where $u_{2k}$ is the coefficient of the tree level superpotential
(\ref{SOtree}). If the tree level superpotential is even function, then
$f(x)$ is also even function. This means that the heperelliptic curve
obtained from this matrix model has $Z_2$ identification and equivalent
to the geometry (\ref{SOgeometry}) obtained in the reduction of Calabi-Yau geometry.

We will consider the matrix models 
with the simplest potential
\begin{eqnarray}
W_{\rm{tree}}=\frac{1}{2}m \Phi^2+\frac{1}{4}g \Phi^4 .
\end{eqnarray}
We can calculate this genus zero free energy in the same way as in the case of $U(N)$ gauge theory along the derivation in \cite{Zuber}. 
Since the eigenvalues $\lambda_i$ of $\Phi$ become continuous in the
large $N$ limit, let us introduce a function  $\lambda(x)$ such that
$\lambda_k=\lambda(k/N)$ for $SO(N)$ and $\lambda_k=\lambda(2k/N)$ for $Sp(N)$.
For real symmertic matric model and self-dual quaternionic matrix model,
the free energies become same and they are given by the integrals
\begin{eqnarray}
\mathcal{F}_0=-\frac{g_s^2 N^2}{g_s}\int_0^1dx \left[\frac{m}{2N}\lambda^2(x)+\frac{g}{4N}\lambda^4(x) \right] +\frac{g_s^2N^2}{2}\int_0^1 \int_0^1 dx dy \log |\lambda(x)-\lambda(y)|.
\end{eqnarray}
We introduce the density of eigenvalues $\rho(\lambda)$ defined as
\begin{eqnarray}
\frac{dx}{d\lambda}=\rho(\lambda),\ \ \ \ \int_{-2a}^{2a}d\lambda \, \rho(\lambda)=1.
\end{eqnarray}
As mentioned above, the geometry derived the matrix models with even function are symmetric. Then we put the interval as symmetric. The equation of motion is rewritten in the large $N$ limit as
\begin{eqnarray}
\frac{m}{g_sN}\lambda +\frac{g}{g_sN}\lambda^3=\int_{-2a}^{2a} \!\!\!\!\!\!\!\!\!\!\!\!- \ \ \ d\mu \frac{\rho(\mu)}{\lambda-\mu}
\end{eqnarray}
where the integral means the principal part. The solution is obtained by introducing the new analytic function,
\begin{eqnarray}
\omega(\lambda)=\int_{-2a}^{2a}d\mu \frac{\rho(\mu)}{\lambda-\mu},
\end{eqnarray}
defined for complex $\lambda$ outside the real interval $(-2a,2a)$. It behaves as $1/\lambda$ when $|\lambda|$ goes to infinity. When $\lambda$ approaches the interval $(-2a,2a)$ it behaves as
\begin{eqnarray}
\omega(\lambda\pm i0)=\frac{m}{g_sN}\lambda +\frac{g}{g_sN}\lambda^3 \mp i\pi \rho(\lambda).
\end{eqnarray}
There is a unique function, which satisfies these requirements 
\begin{eqnarray}
\omega(\lambda)=\frac{m}{g_sN}\lambda +\frac{g}{g_sN}\lambda^3-\left(\frac{m}{g_sN}+2\frac{g}{g_sN}a^2+\frac{g}{g_sN}\lambda^2 \right)\sqrt{\lambda^2-4a^2}
\end{eqnarray}
with the constraint 
\begin{eqnarray}
6ga^4+2ma^2=g_sN. \label{SOconst2}
\end{eqnarray}
Since only the singular part contributes to the density of eigenvalues, we get explicit function for the density
\begin{eqnarray}
\rho(\lambda)&=&\frac{1}{2\pi i}\left(\omega(\lambda+i0)-\omega(\lambda-i0) \right) \nn \\
&=&  \frac{1}{\pi}\left(\frac{m}{g_sN}+2\frac{g}{g_sN}a^2+\frac{g}{g_sN}\lambda^2 \right)\sqrt{\lambda^2-4a^2}. \label{SOsol}
\end{eqnarray}
In terms of these quantities we obtain
\begin{eqnarray}
\mathcal{F}_0&=&-\frac{g_s^2 N^2}{g_s}\int_{-2a}^{2a} d \lambda \, \rho(\lambda) \left[\frac{m}{2N}\lambda^2(x)+\frac{g}{4N}\lambda^4(x) \right] \nn \\ 
{}&&\qquad \qquad + \frac{g_s^2N^2}{2}\int_{-2a}^{2a}\int_{-2a}^{2a} d\lambda \, d\mu\, \rho(\lambda) \rho(\mu) \log |\lambda-\mu| \nn \\
&=& -\frac{S^2}{48}(b^2-1)(9-b^2)+\frac{S^2}{4}\log b^2.
\end{eqnarray}
where $b^2=\frac{2m}{g_sN}a^2$. This is the exact result of genus zero free energy.  
Using the relation (\ref{SOconst2}) we get the genus zero free energy \cite{Zuber}
\begin{eqnarray}
\mathcal{F}_0(g_s)=-\frac{(g_s N)^2}{2} \sum_{k=1}^{\infty} \left(-\frac{3g g_s N}{2m^2} \right)^k \frac{(2k-1)!}{k!(k+2)!}. \label{SOresult}
\end{eqnarray}
For the later discussion for Dijkgraaf-Vafa's conjecture we also calculate the derivation of the genus zero free energy with respect to 't Hooft coupling $S\equiv g_sN$,
\begin{eqnarray}
\frac{\partial \mathcal{F}_0}{\partial S}=-\frac{1}{24}\left[S(b^2-1)(9-b^2)+S^2\left(\frac{\partial b^2}{\partial S}\right)\frac{(b^2-3)(2-b^2)}{b^2} \right]+\frac{S}{2}\log b^2.
\end{eqnarray}
Using the derivative of constraint (\ref{SOconst2}) with respect to $S$
\begin{eqnarray}
\frac{\partial b^2}{\partial S}=-\frac{3gb^4}{2m^2+6gSb^2},
\end{eqnarray}
and substituting the constraint (\ref{SOconst2}) we obtain
\begin{eqnarray}
\frac{\partial \mathcal{F}_0}{\partial S}=-\frac{S}{4}(b^2-1)+\frac{S}{2}\log b^2.
\end{eqnarray}
Since $\mu^2=a^2=\frac{S}{2g\Delta}b^2$,
this result agree with the dual period integral $\Pi$ (\ref{SOdualperi})
in the previous subsection.

\subsection{The check of Dijkgraaf-Vafa conjecture for $SO(N)/Sp(N)$ case}

In \cite{DV3} the Dijkgraaf-Vafa's conjecture is generalized to the
unoriented case. 
With the result (\ref{SOresult}) we can also write down the effective
superpotential for $SO(N)/Sp(N)$ gauge group \footnote{In \cite{DV3} the
glueball superfield was identified with $g_s (N\mp 2)$. Since we have
considered the large $N$ limit, we can ignored the $\mp 2$. Then we identified the glueball superfield with $g_sN$. }
 \begin{eqnarray}
 W_{\mathrm{eff}}(S)&=&\hat{N}S \log \left(\frac{S}{\Lambda_0^3}\right)-2\pi i \tau S+\hat{N} \frac{\partial \mathcal{F}_0}{\partial S} \nn \\
 {}&=&\hat{N}S \log \left(\frac{S}{\Lambda_0^3}\right)-2\pi i \tau S+\hat{N} \left[-\frac{S}{4}(b^2-1)+\frac{S}{2}\log b^2 \right]
 \end{eqnarray}
where $\hat{N}=N-2$ for $SO(N)$ and where $\hat{N}=N+2$ for $Sp(N)$. 
 This effective superpotential  agree exactly  with the result (\ref{SOCYexact}) from the Calabi-Yau manifold with fluxes.
 \begin{eqnarray}
W_{\mathrm{eff}}(S) \!\! &=& \!\!  \hat{N}S \log \left(\frac{S}{\Lambda_0^3}\right)-2\pi i \tau S-\hat{N} \sum_{k=1}^{\infty} \left(-\frac{3g }{2m^2} \right)^k \frac{S^{k+1}}{2} \frac{(2k-1)!}{k!(k+1)!} \nn \\{} \!\! &=& \!\!  \cdots+\hat{N} \left[\frac{3}{2}\left(\frac{g}{4m^2} \right)S^2-\frac{9}{2} \left(\frac{g^2}{8m^4} \right)S^3+\frac{45}{2} \left(\frac{g^3}{16m^6} \right)S^4+\cdots \right], \label{SOeff2}
\end{eqnarray} 
This is  the exact effective superpotential summed
up all instanton corrections for the $SO(N)/Sp(N)$ gauge theories.  
As already mensioned this model is captured by $\Phi^4$-theory
with single line in which non-planar d
iagrams are omitted. 
The diagrams in Fig.1 give the contribution to the second term of (\ref{SOeff}).
\begin{figure}[htbp]
\begin{center}
\includegraphics[width=12cm,height=3cm]
{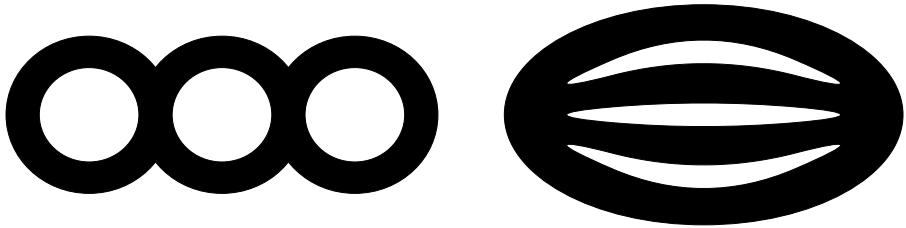}
\caption{3-loop diagrams contributing to the fourth order of $S$
}
 \label{planar}
\end{center}
\end{figure}
This result also 
agree perfectly with (\ref{SOeff}). Thus Dijkgraaf-Vafa's proposal for 
the unoriented case is
explicitly checked for the perturbation around
$\Phi=0$.

\noindent
{\large{\bf Acknowledgements}}

The authors would like to thank Norisuke Sakai and Katsushi Ito for
useful suggestions.
The authors are also obliged to Yoshiyuki Watabiki for stimulating
discussions.
H.F. is supported by JSPS research fellowship 
for young scientists.


\begin{thebibliography}{100}
\bibitem{Hooft}
  G. 't Hooft, ``A Planar Diagram Theory For Strong Interactions,'' Nucl. Phys. \textbf{B72} (1974) 461.
\bibitem{GV}
  R. Gopakumar and C. Vafa, ``On the gauge theory/geometry correspondence,'' Adv. Theor. Math. Phys. \textbf{3} (1999) 1415.
\bibitem{SV}
  S. Sinha and C. Vafa, ``SO and Sp Chern-Simons at large N,'' hep-th/0012136.
\bibitem{OV}
  H. Ooguri and C. Vafa, ``Worldsheet derivation of a large N duality,'' Nucl. Phys. \textbf{B641} (2002) 3.
\bibitem{AMV}
  M. Aganagic, M. Marino and C. Vafa, ``All loop topological string amplitudes from Chern-Simons theory,'' hep-th/0206164. 
\bibitem{VAFA}
  C. Vafa, ``Superstrings and topological strings at large N,'' J.Math. Phys. \textbf{42} (2001) 2798.
\bibitem{CIV}
  F. Cachazo, K. Intriligator and C. Vafa, ``A large N duality via geometric transition,'' Nucl. Phys. \textbf{B603} (2001) 3.
  \bibitem{CKV}
  F. Cachazo, S. Katz and C. Vafa, ``Geometric transitions and N=1 quiver theories,'' hep-th/0108120.
  \bibitem{CFIKV}
  F. Cachazo, B. Fiol, K. Intriligator, S. Katz and C. Vafa, ``A geometric unification of dualities,'' Nucl.Phys. \textbf{B628} (2002) 3.
\bibitem{OH}
  J.D. Edelstein, K. Oh and R. Tatar, ``Orientifold, geometric transition
	and large N duality for SO/Sp gauge theories,''
	JHEP.\textbf{0105} (2001) 9.
\bibitem{FO}
  H. Fuji and Y. Ookouchi, ``Confining phase superpotentials for SO/Sp gauge theories via geometric transition,'' hep-th/0205301.
\bibitem{DOT}
  K. Dasgupta, K. Oh and R. Tatar, ``Geometric transition, large N
	dualities and MQCD,'' Nucl. Phys. \textbf{B610} (2001) 331,
	K. Dasgupta, K. Oh and R. Tatar, ``Open/closed string dualities
	and Seiberg duality from geometric transitions in M-theory,''
	hep-th/0106040, K. Dasgupta, K. Oh and R. Tatar, ``Geometric transition versus cascading solution,'' JHEP \textbf{0201} (2002) 31.
\bibitem{CV}
  F. Cachazo and C. Vafa, ``N=1 and N=2 geometry from fluxes,'' hep-th/0206017.
\bibitem{Gukov}
  S. Gukov, C. Vafa and E. Witten, ``CFT's From Calabi-Yau Four-folds,'' Nucl.Phys. B584 (2000) 69, 
 S. Gukov, ``Solitons, Superpotentials and Calibrations,'' Nucl.Phys. B574 (2000) 169.
\bibitem{TV}
  T.R. Taylor and C. Vafa, ``RR flux on Calabi-Yau and partial supersymmetry breaking,'' Phys.Lett. \textbf{474} (2000) 130.



  
\bibitem{DV1}
  R. Dijkgraaf and C. Vafa, ``Matrix models, topological strings, and supersymmetric gauge theories,'' hep-th/0206255.
 \bibitem{DV2}
  R. Dijkgraaf and C. Vafa, ``On geometry and matrix models,'' hep-th/0207106.
 \bibitem{DV3}
  R. Dijkgraaf and C. Vafa, ``A perturbative window into non-perturbative physics,'' hep-th/0208048.

\bibitem{CM}
L.Chekhov and A.Mironov, ``Matrix models vs. Seiberg-Witten/Whitham theories,'' hep-th/0209085

\bibitem{Dorey1}
  N. Dorey, T.J. Hollowood, S.P. Kumar and A. Sinkovics, ``Exact superpotentials from matrix models,'' hep-th/0209089.
\bibitem{Dorey2}
  N. Dorey, T.J. Hollowood, S.P. Kumar and A. Sinkovics, ``Massive vacua of N=1* theory and S-duality from matrix models,'' hep-th/0209099.

\bibitem{Fe}
F. Ferrari,
``On exact superpotentials in confining vacua,'' hep-th/0210135.
\bibitem{Zuber}
  E. Br\'ezin, C. Itzykson, G. Parisi, and J.B. Zuber, ``Planar diagrams,'' Commun. Math. Phys. \textbf{59} (1978) 35.

\bibitem{VY}
  G. Veneziano and S. Yankielowicz, ``An effective Lagrangian for the pure N=1 supersymmetric Yang-Mills theory,'' Phys. Lett. \textbf{B113} (1982) 231.
\bibitem{Mor}
  A. Morozov, ``Matrix Models as Integrable Systems,'' hep-th/9502091.

\bibitem{Me}
 M.L. Mehta, ``Random Matrices,''  (Academic Press, New York, 1967).

\bibitem{FGZ}
  P.Di. Francesco, P. Ginsparg and J. Zinn-Justin, ``2-D gravity and random matrices,'' Phys.Rept.\textbf{254} (1995) 1.\bibitem{BEH}
  M. Bertola, B. Eynard and J. Harnad, ``Partition functions for matrix models and isomonodromic tau functions,'' nlin.SI/0204054.

\end{thebibliography}
\end{document}